\documentclass[10pt]{{elsarticle}}

\usepackage[a4paper, total={6in, 8in}]{geometry}
\usepackage[utf8]{inputenc}
\usepackage{amsfonts}
\usepackage{amssymb}
\usepackage{amsthm}
\usepackage{amsmath}

\usepackage{mathrsfs}
\usepackage{graphicx}
\usepackage{esvect}
\usepackage{tikz}
\usepackage{verbatim}
\usepackage{bbold}
\usepackage{caption}
\usepackage{acronym}
\usepackage{overpic}
\usepackage{amsmath}
\usepackage{xcolor}
\usepackage{sectsty}
\usepackage{changepage}
\usepackage[procnumbered, ruled]{algorithm2e}
\DontPrintSemicolon
\usepackage[mathscr]{euscript}
\usepackage[utf8]{inputenc}
\usepackage[english]{babel}
\usepackage{subcaption}
\usepackage{mwe}
\LinesNumbered
\sectionfont{\normalsize}

\newtheorem{theorem}{Theorem}[subsection]
\newtheorem{corollary}{Corollary}[subsection]
\newtheorem{definition}{Definition}[subsection]

\newtheorem{lemma}{Lemma}[subsection]

\newcommand{\rvline}{\hspace*{-\arraycolsep}\vline\hspace*{-\arraycolsep}}
\newcommand{\bigzero}{\mbox{\normalfont\Large 0}}
\newcommand{\Medzero}{\mbox{\normalfont 0}}

\newcommand{\hidden}[1]{}
\newcommand{\N}{\mathbb{N}}

\newcommand{\R}{\mathbb{R}}

\newtheorem{example}{Example}[subsection]

\acrodef{DFT}[DFT]{Discrete Fourier Transform}
\acrodef{FST}[FST]{Graph Fourier Transform}
\acrodef{EP}[EP]{Equitable Partition}
\acrodef{ETP}[ETP]{Equitable Transmitting Partition}
\acrodef{ERP}[ERP]{Equitable Receiving Partition}
\acrodef{LEP}[LEP]{Local Equitable Parition}
\acrodef{CEP}[CEP]{Complete Equitable Partitions}

\makeatletter
\renewcommand*\env@matrix[1][*\c@MaxMatrixCols c]{%
	\hskip -\arraycolsep
	\let\@ifnextchar\new@ifnextchar
	\array{#1}}
\makeatother

\begin{document}
	\begin{frontmatter}
		\date{\today}
		
		\title{Minimal Specialization: Coevolution of Network Structure and Dynamics}
		\author[King]{Annika King}
		\author[Smith]{Dallas Smith}
		\author[Webb]{Benjamin Webb}
		\address[King]{Department of Mathematics, Brigham Young University, Provo, UT 84602, USA, aking@mathematics.byu.edu}
		\address[Smith]{Department of Mathematics, Utah Valley University, Orem, UT 84058, USA, dallas.smith@uvu.edu}
		\address[Webb]{Department of Mathematics, Brigham Young University, Provo, UT 84602, USA, bwebb@mathematics.byu.edu}

		
		\begin{abstract}
         The changing topology of a network is driven by the need to maintain or optimize network function. As this function is often related to moving quantities such as traffic, information, etc. efficiently through the network the structure of the network and the dynamics on the network directly depend on the other. To model this interplay of network structure and dynamics we use the \emph{dynamics on} the network, or the dynamical processes the network models, to influence the \emph{dynamics of} the network structure, i.e., to determine where and when to modify the network structure. We model the dynamics on the network using Jackson network dynamics and the dynamics of the network structure using  \emph{minimal specialization}, a variant of the more general network growth model known as \emph{specialization}. The resulting model, which we refer to as the \emph{integrated specialization model}, coevolves both the structure and the dynamics of the network. We show this model produces networks with real-world properties, such as right-skewed degree distributions, sparsity, the small-world property, and non-trivial equitable partitions. Additionally, when compared to other growth models, the integrated specialization model creates networks with small diameter, minimizing distances across the network. Along with producing these structural features, this model also sequentially removes the network's largest bottlenecks. The result are networks that have both dynamic and structural features that allow quantities to more efficiently move through the network.
		\end{abstract}
		
		
	\end{frontmatter}

 \section{Introduction}
Networks studied in the biological, social, and technological sciences are inherently dynamic in that the state of the  network's components evolve overtime. Technological and traffic networks show phase-transition type dynamics \cite{DynamicNetworks,TechTraffic}, gene regulatory networks experience boolean dynamics \cite{DynamicNetworks,bioinformatics}, metabolic networks exhibit flux-balance dynamics \cite{DynamicNetworks,metabolic}, and the human brain has been shown to have synchronous and other dynamic behavior \cite{DynamicNetworks,glass2001synchronization}.

The first type of dynamics is most often referred to as the \emph{dynamics on the network}, referring to the changing states of the network's components. The second type of dynamics, which is the evolving topology of the network, is referred to as the \emph{dynamics of the network}. To a large extent the study of network dynamics focuses on one of these two types of dynamics, meaning either the network's structure is fixed and the dynamics on the network are studied, or the dynamics on the network are ignored and the evolving structure of the network is studied. However, in real-world networks, these two types of dynamics typically influence one another \cite{bullmore2009complex,barabasi2004network,ash1995dynamic}. For instance, as traffic increases in a traffic network, the Internet, a supply chain, etc., new routes are added, which in turn creates new traffic patterns.


Here, we consider the interplay between these two types of dynamics. Specifically, we introduce a model that uses the dynamics on the network to determine where to evolve the network's structure. This change in structure in turn alters the dynamics on the network, and this coevolving back and forth of structure and dynamics is analyzed. For the dynamics on the network, we use a Jackson network model, which describes queuing systems, e.g. transportation networks, website traffic, etc. \cite{Jackson}. Under certain conditions, Jackson networks have a \emph{stationary distribution}, which can be thought of as a \emph{globally attracting fixed point} that the dynamics on the network tend to. This allows us to determine asymptotically high-stress areas, or areas of maximal load, within the network \cite{stationary}. These areas with maximal load are where we perform the structural evolution of the network via minimal specialization, which maintains the function of the network while separating the number of tasks, i.e. load, of these high-stress areas.

Specialization, as a mechanism of growth, is a phenomenon observed in  many real-world networks including biological \cite{sporns2013network,sole2008spontaneous,espinosa2010specialization}, social \cite{salz2001development}, and airline hubs in transportation networks \cite{burghouwt2014long}. Specialization allows a network to perform increasingly complex tasks by copying parts of the network and dividing the original network connections between these copies. The resulting specialized network maintains the functionality of the network by preserving all the network paths so that the ability to route information, etc. is maintained. Recently, network growth models for specialization have been studied in \cite{Specialization, Spectral, Synchronization}, where the authors describe how specialization creates real-world properties in a network over time \cite{Specialization}, maintains intrinsic stability \cite{Spectral}, and creates synchronous dynamics \cite{Synchronization}. 

The specific model we propose uses what we refer to as \emph{minimal specialization} to evolve the structure of the Jackson network in the area of highest maximal load (see Sections \ref{Weights} and \ref{interplay}). As this method incorporates both structure and dynamics we refer to it as the \emph{integrated specialization model}. Here, we show that the integrated specialization model creates networks that have structures observed in real-world networks, including right-skewed degree distributions, sparsity, and the small-world property. Moreover, when compared to other growth models, the integrated specialization model creates networks with small diameter, i.e., networks that have relatively small distances across the network. Aside from these structural features, the integrated specialization model sequentially removes the network's largest bottlenecks as the network's structure evolves. This results in networks whose structure and dynamics are both increasingly well-adapted to allow quantities to move efficiently through the network.


The paper is structured as follows. Section \ref{background} introduces basic notation, Jackson networks, and the equations used to model the dynamics on Jackson networks. In Section \ref{Weights}, we  motivate and introduce the concept of \emph{minimal specialization}, showing that certain topological and dynamical properties of the original network are maintained under these operations. In Section \ref{interplay}, we use the dynamics, i.e., the stationary distribution, on a Jackson network to determine where to perform minimal specialization of the network. We then extend this method, which we refer to as \emph{minimal dynamic specialization}, to more general networks using the network's \emph{eigenvector centrality} to specialize the network structure. In Section \ref{properties}, we show that growing a Jackson network repeatedly using minimal dynamic specialization, referred to as the \emph{integrated specialization model}, creates structural properties observed in real-world networks. We also show that the maximum eigenvector centrality of the network decreases meaning the integrated model sequentially reduces areas of high stress in the network, on average. Section \ref{comapre_models} compares the integrated specialization model to other growth models, where we show that the integrated specialization model is comparable in creating real-world properties while being more efficient in creating small diameter networks and networks with equidistributed traffic loads. Section \ref{ep ch} introduces equitable partitions and shows that minimal specialization creates and preserves non-trivial equitable partition elements, which are related to symmetries in the network and are common structures observed in real-world networks.

\section{Background}\label{background}

Real-world networks perform specific functions. The underlying structure of a network, represented by a graph, is key to its performance (\cite{stelling2002metabolic,sandstrom2008performance,Structure}). Formally, this is given by the graph $G = (V,E,W)$, where $V$ is a \emph{vertex set} (or node set) and $E$ is an \emph{edge set}. The vertices in $V$ represent the network components, or objects, while the edges in $E$ represent the connections or interactions between these objects. We let $V = \{1,2, \dots , n \}$ with $i$ representing the $ith$ component of the network. An edge from $i$ to $j$, denoted $e_{ij}$ is used to represent the $ith$ node affecting the $jth$ node. Here, the graphs we consider are directed graphs, meaning edges of the graph are directed, noting that undirected graphs can be viewed as a special case of directed graphs. The function $W: E \to \R$ gives the edge weight of each $e_{ij} \in E$, where $W(e_{ij})$ can represent the strength of the interaction, etc. If $G$ is an unweighted graph, then $W \equiv 1$ and we write $G = (V,E)$.  

The underlying graph structure $G = (V,E,W)$ of a network can alternatively be represented by its \emph{adjacency matrix} $A = A(G) \in \R^{n \times n}$, with entries
\[A_{ij} = \begin{cases}
  W(e_{ji})  &  \text{ if }  e_{ji} \in E\\
  0 & \text{else.}
\end{cases}\]
We note that this orientation corresponds to right multiplication by a column vector. We let $\sigma(A)$ denote the \emph{eigenvalues} of the matrix $A$, and $\rho(A) = \max\{|\lambda| : \lambda \in \sigma(A)\}$ the \emph{spectral radius} of $A$. Since there is a one-to-one relationship between a graph and its adjacency matrix, we will use the two interchangeably.

To study the interplay between the dynamics on and of a network, we consider Jackson networks, which are used to model queuing systems \cite{Jackson}. The flow of a given quantity, e.g. traffic, information, etc., through a Jackson network is modeled as the discrete-time affine dynamical system 
\[\mathbf{x}^{(k+1)} = A\mathbf{x}^{(k)} + \mathbf{\gamma}\]
where $\mathbf{x}^{(k)} \in \R^n$ is the \emph{state} of the network giving the state $x_i^{(k)}$ of each component at time $k \geq 0$. The graph $G = (V,E,W)$ associated with the Jackson network is the graph with the adjacency matrix $A \in [0,1]^{n \times n}$, called the system's \emph{transition matrix} where $A_{ij} \in [0,1]$ is the probability of transitioning from vertex $i$ to vertex $j$. The vector $\mathbf{\gamma} \in \R^n$ gives the \emph{external inputs} to the system. A Jackson network can also have \emph{internal loss}, meaning there is a probability of information, traffic, etc. leaving the system. This occurs when the probability of transitioning from node $i$ to any other node is less than 1, i.e., the $ith$ column of the transition matrix sums to less than 1. For simplicity, we assume there is no external input or internal loss, meaning $\mathbf{\gamma} \equiv \mathbf{0}$ and $A$ is column stochastic. Thus,
\[\mathbf{x}^{(k+1)} = A\mathbf{x}^{(k)} = A^k \mathbf{x}^{(0)} \label{system} \tag{$1$}\]
for the \emph{initial vector} of quantities $\mathbf{x}^{(0)} \in \R^n$. This configuration induces the discrete-time linear dynamical system $\left(A, \R^{n}\right)$ where the matrix $A = A(G)$ is the adjacency matrix of the graph $G = (V,E,W)$.

The asymptotic behavior of linear systems, such as the Jackson networks $\left(A, \R^n\right)$, are fairly well understood. If $A$ is primitive, then the \emph{spectral radius} is an algebraically simple eigenvalue of $A$ and $\lim_{k \to \infty} (\rho^{-1}A)^k = \mathbf{x}\mathbf{y}^\intercal$ where $\mathbf{x}$ and $\mathbf{y}$ are the right and left leading eigenvectors associated with $\rho$, respectively, with $\|\mathbf{x}\|_1 = 1$ and $\mathbf{x}^\intercal \mathbf{y} = 1$. Moreover, since $A$ is stochastic, we have $\rho = 1$ and $\mathbf{y} = \mathbb{1}$. That is, for the initial condition $\mathbf{x}^{(0)} \geq 0$, we have \[\lim_{k \to \infty} A^k \mathbf{x}^{(0)} = \mathbf{x}\mathbb{1}^\intercal \mathbf{x}^{(0)} = \|\mathbf{x}^{(0)}\|_1\mathbf{x}. \label{dynamics} \tag{$3$}\]
Thus the sum of the quantities moving through the network is constant in time \cite{horn2012matrix}. Additionally the \emph{asymptotic state} of the system is a scaled version of the leading eigenvector $\mathbf{x}$, which when scaled to one, is referred to as the systems \emph{stationary distribution} \cite{stationary}. For a Jackson network $(A,\R^n)$ with primitive matrix $A$, the stationary distribution is a \emph{globally attracting fixed point} of the system (see Equation \ref{dynamics}). This stationary distribution indicates the long-term dynamics of the Jackson network $\left(A,\R^n\right)$. Specifically, it tells us which nodes, on average, carry the highest amount of information, traffic, stress, etc. The main idea behind the model we propose is to use this globally attracting fixed point to determine where the long-term, high-stress areas of the network are (see Section \ref{interplay}). To alleviate this stress, we modify the structure of the network accordingly, using the notion of \emph{minimal specialization}. 

\section{Topological Network Dynamics: \emph{Minimal Specialization}}\label{Weights}

In order to model the interplay between the dynamics on and the dynamics of the network, we need to determine \emph{how} to evolve the structure of the network. As described in the introduction, network specialization has been observed in a number of real-world networks \cite{sporns2013network,sole2008spontaneous,espinosa2010specialization,salz2001development,burghouwt2014long}, and specialization models have been the focus of a number of recent papers \cite{Specialization, Spectral, Synchronization}.
In this work, we explore two main deviations from these specialization models. The first is related to the observation that most real-world specializations occur at small scales. For example, if a transportation route experiences high use, usually the route is only modified at its point of highest traffic. To reflect this, our model of minimal specialization adds the fewest number of nodes and edges required to modify the flow through the network while maintaining functionality.

\begin{definition}[\textbf{Minimal Specialization}]\label{min_specialization}
    For the graph $G = (V,E,W)$ with $|V| > 1$, let $i \in V$ such that $i$ has at least two outgoing edges. Let $e_{ij} \in E$ with $i\neq j$ and let $w = W(e_{ij})$. Let $\overline{G} = \left(\overline{V},\overline{E},\overline{W}\right)$ be the graph where\\
    \indent (i) $\overline{V} =  V \cup   \{\overline{i}\}$;\\
    \indent (ii) $\overline{E} = \left(E \setminus \{e_{ij}\}\right) \cup \{e_{{\overline i}j}\}\ \cup \{e_{k \overline{i}} \; |\; \exists \; e_{ki} \in E \}$; and\\
    \indent (iii) $\overline{W}(e_{\alpha\beta})=\begin{cases}
        1 \quad &\text{if} \quad \alpha=\overline{i}, \ \beta=j\\
        (1-w)^{-1}W(e_{\alpha i}) \quad &\text{if} \quad \alpha=i,\ \beta\neq j\\
        (1-w)W(e_{\alpha\beta}) \quad &\text{if} \quad \alpha\neq i, \ \beta=i\\
        wW(e_{\alpha i}) \quad &\text{if} \quad \alpha\neq i, \ \beta=\overline{i}. \\
        W(e_{ii}) \quad &\text{if} \quad \alpha = i, \beta = i \\
        w(1-w)^{-1} W(e_{ii}) \quad &\text{if} \quad \alpha = i, \beta = \overline{i} \\
        W(e_{\alpha \beta})\quad &\text{else}.
          \end{cases}$\\
    We refer to the graph $\overline{G} = \left(\overline{V},\overline{E},\overline{W}\right)$ associated with the system $\left(\overline{A},\R^{(n+1)}\right)$ as the \emph{minimal specialization} of the graph $G$ over $i \in V$ with edge $e_{ij} \in E$.
    
\end{definition} 

In Definition \ref{min_specialization}, $\overline{i} \in \overline{V}$ can be thought of as a copy of $i \in V$, whose specialized function is to maintain the weighted connections to $j \in V$ that were previously maintained by $i$ through $e_{ij}$. This results in vertices specialized into two parts, where $i$ performs its previous task with the exception of routing traffic to $j$, which is now executed by its specialized copy $\overline{i}$ (see Example \ref{Ex_Min_Spec}).

It is important  that $i\in V$ has at least two out-edges in order to be specialized. If it has only one out-edge, then, after specialization, it would have no out-going edges, becoming a \emph{dangling node}, essentially meaning that information, information, etc., gets trapped at that node. 

We note that the edge weight update in Definition \ref{min_specialization} maintains the stochastic nature of the Jackson network (see Section \ref{main_stability}). This manner of updating could be done for any $0 < w < 1$ and still produce a stochastic system, but choosing $w = W(e_{ij})$ maintains the proportion of network quantities passing through $i$ and its copy $\overline{i}$ after specialization that previously passed through $i$ before specialization.

\subsection{Topological and Dynamical Properties of Minimal Specialization}\label{main_stability}
Minimal specialization induces a new system $(\overline{A},\R^{n+1})$ with new dynamics. Here we show this new system is also a Jackson network that inherits the structural and stability properties of the original Jackson network $\left(A, \R^n \right)$ under mild conditions. In particular, we will show that if $A$ is primitive then $\overline{A}$ is primitive, so that the specialized Jackson network $(\overline{A},\R^{n+1})$ has a stationary distribution $\overline{\mathbf{x}} \in \R^{n+1}$. To prove this requires the following two lemmata.

\begin{lemma}[\textbf{Preservation of a Strongly Connected Graph}]\label{strong}
If $\overline{G} = \left(\overline{V},\overline{E},\overline{W}\right)$ is a minimal specialization of the graph $G = (V,E,W)$ and $G$ is strongly connected, then $\overline{G}$ is strongly connected.
\begin{proof}
    We note that under minimal specialization, the only deleted edge in $G$ is $e_{ij}$, so any path in $G$ that does not contain $e_{ij}$ will also be in $\overline{G}$. Thus we only need to verify three things: (1) there is a path in $\overline{G}$ from $i$ to $j$. Thus, any path in $G$ that uses $e_{ij}$ now uses the path from $i$ to $j$, ignoring any cycles created. (2) There are paths in $\overline{G}$ from any node to $\overline{i}$ and (3) from $\overline{i}$ to any node.
    
    First, we note that by constraints in definition \ref{min_specialization}, $i \in V$ has an edge $e_{ik} \in E$ with $k \neq j$. Moreover, since $G$ is strongly connected, there is at least one edge $e_{hi} \in E$. Thus, there is a path from $k$ to $h$ and we have $e_{h\overline{i}} \in \overline{E}$. Therefore, we have the path $ i \rightarrow k \rightarrow \cdots \rightarrow h \rightarrow \overline{i} \rightarrow j$.

    Second, there is a path in $G$ from any node to $i$ and this path does not contain $e_{ij}$ since it would create a cycle. Thus, we can take the penultimate node in that path and traverse the edge from that node to $\overline{i}$, and thus we have a path in $\overline{G}$ ending at $\overline{i}$.
    
    Finally, it is easy to see there is a path from $\overline{i}$ to any node, since we have the edge $e_{\overline{i}j}$, and there is a path from $j$ to any node.
\end{proof}
\end{lemma}


\begin{lemma}[\textbf{Spectral Evolution of Jackson Networks}]
\label{Eigenvector}
Suppose $\overline{G} = \left(\overline{V},\overline{E},\overline{W}\right)$ is the minimal specialization of $G = (V,E,W)$ over $i \in V$ with edge $e_{ij} \in E$. If $w = W(e_{ij})$ and $(\lambda, \mathbf{x})$ is an eigenpair of $A = A(G)$ then $(\lambda, \overline{\mathbf{x}})$ is an eigenpair of $\overline{A} = A(\overline{G})$ where, 
$$\overline{x}_{\ell} = \begin{cases}
     (1 - w) x_{\ell}  & \text{  if } \ell = i \\
     w x_{\ell}  & \text{  if } \ell = \overline{i} \\
     x_{\ell}  & \text{  else. } \\
\end{cases}$$ In particular, $x_i = \overline{x}_i + \overline{x}_{\overline{i}}$ so $\|\mathbf{x}\|_1 = \|\mathbf{\overline{x}}\|_1$. The additional eigenvalue of $\overline{A}$ is $0$, meaning $\sigma(\overline{A}) = \sigma(A) \cup \{0\}.$
\\

\begin{proof}
We will first consider how minimal specialization transforms the adjacency matrix. Without loss of generality, we can order our the rows and columns of our matrix so that we have

\begin{align}
A = \left[\begin{array}{c c c c | c | c }
& & A{[\{ji\}^c,\{i\}^c]} & & \rvline & 
\begin{matrix}
 A_{1i}\\
A_{2i} \\
\vdots \\
A_{(j-1)i}
\end{matrix}
 \\
\hline
A_{j1} & A_{j2} & \dots & A_{jj} & \rvline & w\\
A_{i1} & A_{i2} & \dots & A_{ij} & \rvline & A_{ii}  \\
\end{array}\right] \in \R^{n \times n}
\end{align} \label{eq:A}

where $A[\{j,i\}^c, \{i\}^c]$ denotes the matrix $A$ with the $jth$ and $ith$ rows excluded and the $ith$ column excluded.

When we perform minimal specialization, the weight updates found in definition \ref{min_specialization} are reflected in the adjacency matrix as follows:
 
\begin{align} \overline{A} = \left[\begin{array}{c c c c |c| c | c }
&  A{[\{j,i,\overline{i}\}^c,\{i,\overline{i}\}^c]} & & \rvline & 
\begin{matrix}
 \frac{A_{1i}}{(1-w)}\\
\frac{A_{2i}}{(1-w)} \\
\vdots \\
\frac{A_{(j-1)i}}{(1-w)}
\end{matrix}
& \rvline & \bigzero \\
\hline
A_{j1}  & \dots & A_{jj} & \rvline & 0 & \rvline & 1 \\
(1-w)A_{i1}  & \dots & (1-w)A_{ij} & \rvline & A_{ii} & \rvline & 0 \\
w A_{i1}  & \dots & w A_{ij} & \rvline & \frac{wA_{ii}}{1-w} & \rvline & 0
\end{array}\right]  \in \R^{(n+1) \times (n+1)}.
\end{align} \label{eq:A_bar}

Let $\overline{\mathbf{x}}$ be the vector whose entries are given by
$$\overline{x}_{\ell} = \begin{cases}
     (1 - w) x_{\ell}  & \text{  if } \ell = i \\
     w x_{\ell}  & \text{  if } \ell = \overline{i} \\
     x_{\ell}  & \text{  else } \\
\end{cases}$$.

Then for $k \in \{1, \dots j-1\}$, the $kth$ entry of the product $\overline{A}\overline{\mathbf{x}}$ is
\[ \left(\overline{A} \mathbf{\overline{x}}\right)_k = \sum_{s = 1}^j A_{ks} x_s + \frac{A_{ki}}{1-w} (1-w) x_i + 0 \cdot wx_i = \sum_{s = 1}^j A_{ks} x_s + A_{ki} x_i = \left(A\mathbf{x}\right)_k =  \lambda x_k.\] For the $jth$ entry of the product $\overline{A}\overline{\mathbf{x}}$ we have,
\[\left(\overline{A}\mathbf{\overline{x}}\right)_j = \sum_{s = 1}^j A_{js} x_s + 0 + w x_i = \sum_{s = 1}^j A_{js} x_s +A_{ji} x_i = \left(A\mathbf{x}\right)_j =  \lambda x_j. \]
For the $ith$ entry of the product $\overline{A}\overline{\mathbf{x}}$ we have,
\[\left(\overline{A}\overline{\mathbf{x}}\right)_i = \sum_{s = 1}^j (1-w)A_{is} x_s + A_{ii}(1-w)x_i + 0 = (1-w) \sum_{s = 1}^i A_{is} x_s = (1-w) \left(A\mathbf{x}\right)_i = \lambda (1 - w) x_i.\]
Finally, for the $\overline{i}th$ entry of the product $\overline{A}\overline{\mathbf{x}}$ we have,
\[\left(\overline{A}\mathbf{\overline{x}}\right)_{\overline{i}} = \sum_{s = 1}^j w A_{is} x_s + \frac{wA_{ii}}{1-w} (1-w)x_i + 0 = w \sum_{s = 1}^i A_{is} x_s = w \left(A\mathbf{x}\right)_{i} =  \lambda w x_i.\]
Thus we have the product  $\overline{A}\mathbf{\overline{x}} = \lambda \mathbf{\overline{x}}$ and therefore $\mathbf{\overline{x}}$ is an eigenvector with eigenvalue $\lambda$.

Finally, The last two rows of $\overline{A}$ are scalar multiples of each other. Since the last row is a new row, a new linear dependence is created in the matrix, and thus the additional eigenvalue is 0.
\end{proof}
\end{lemma}

\begin{theorem}[\textbf{Preservation of Asymptotic Dynamics}]\label{stable}
    Let $G = (V,E,W)$ and assume $A = A(G) \in [0,1]^{n \times n}$ is primitive and stochastic. Then $\overline{A} = A(\overline{G}) \in [0,1]^{(n+1) \times (n+1)}$, associated with any minimal specialization $\overline{G} = (\overline{V},\overline{E},\overline{W})$, is primitive and stochastic. Therefore, $$\lim_{n \to \infty} \overline{A}^k \overline{\mathbf{x}}^{(0)} = \|\mathbf{x}^{(0)}\|_1 \overline{\mathbf{v}} \ \ \text{for any} \ \ \mathbf{x}^{(0)} \in \R_{\geq 0}^{n+1} $$ where  $\overline{\mathbf{v}}$ is the stationary distribution of $\overline{A}$.
    
    \begin{proof}
    Given that $A$ is stochastic and using Equation \ref{eq:A_bar} in Lemma \ref{Eigenvector}, we see that for $k \in \{1,2, \cdots j\} \cup \{\overline{i}\}$, the $kth$ column of $\overline{A}$ sums to 1. For column $i$, it follows from combining $A_{ii}$ and $w(1-w)^{-1}A_{ii}$ and recognizing that the sum of the $ith$ column of $A$ excluding  $A_{ji}$ is $1-w$. Thus, $\overline{A}$ is stochastic.
    
    From Lemma \ref{strong}, we have that $\overline{G}$ is strongly connected. From Lemma \ref{Pres}, we have $\sigma(\overline{A}) = \sigma(A) \cup \{0\}$. Since $A$ is primitive, $\rho(A) > 0$ is the only eigenvalue of maximum modulus. Thus $\rho(\overline{A}) = \rho(A)$ is the only eigenvalue of $\overline{A}$ of maximal modulus. Thus $\overline{A}$ is primitive (see \cite{horn2012matrix} for the definition of primitive and equivalent characterizations). The dynamic consequences now follow from arguments in Section \ref{background}.
    \end{proof}
\end{theorem}

\begin{figure}[h!]
     \centering
     \caption{Example of Minimal Specialization}
     \begin{subfigure}[t]{.8\textwidth}
        \vskip 0pt
         \includegraphics[width=\textwidth]{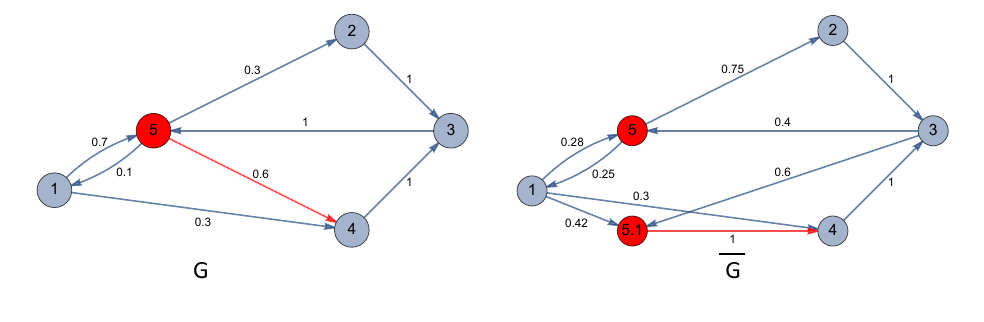}
    \caption{An example of \textit{minimal specialization} where the graph $\overline{G}$ on the right is a specialized version of the graph $G$ on the left. Here, node 5 with edge $e_{54}$ is specialized resulting in the node (red on left). Via specialization, an edge from node $5.1$ to node $4$ is added (red on right). Nodes $1$ and $3$, which have an edge into node $5$ now also have edges into node $5.1$. Additionally, the edge from node $5$ to node $4$ is deleted but all other outgoing edges of 5 remain. The weights are updated according to the given scheme (see Definition \ref{min_specialization}).}
      \label{fig:SpecEx}
     \end{subfigure}
      \vskip\baselineskip
     \begin{subfigure}[t]{.8\textwidth}
        \vskip 0pt
\includegraphics[width=\textwidth]{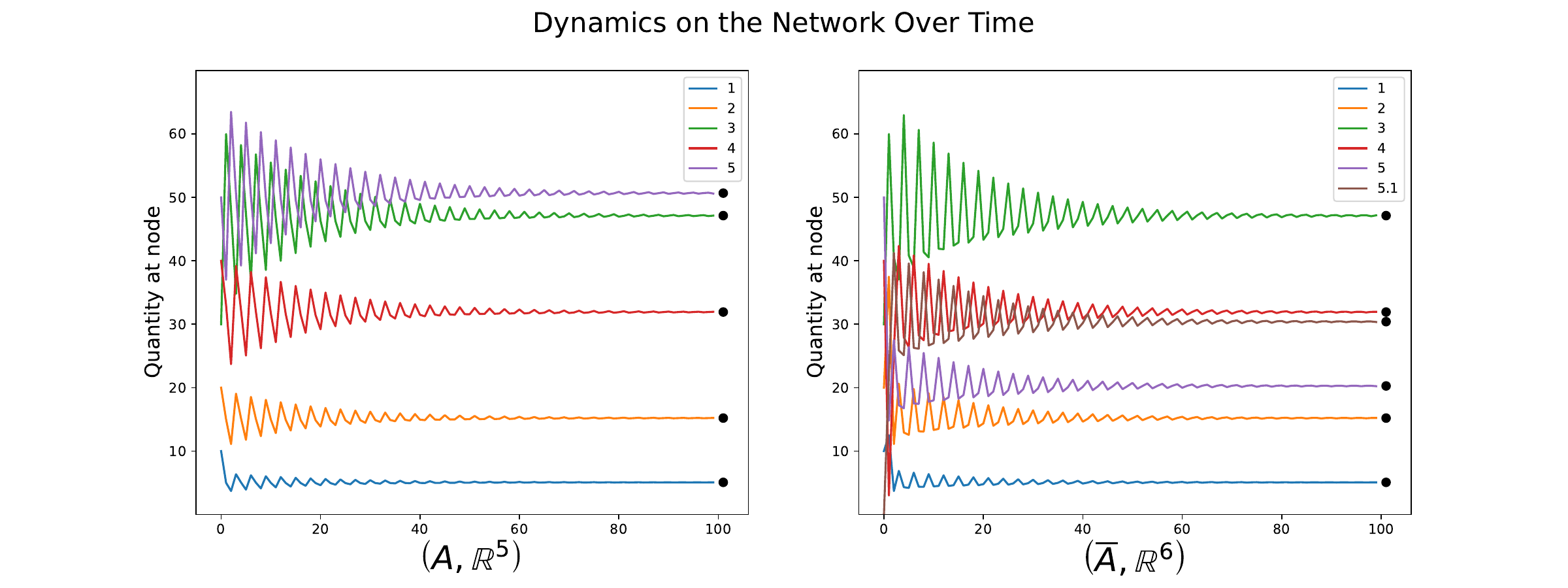}
    \caption{The dynamics on the network over time, for the network in Figure \ref{fig:SpecEx}. For the initial network,the system's dynamics are shown for the initial condition $\mathbf{x}_0 = [10,20,30,40,50]$ (left). Here, the long-term behavior at each node tend to a scaled version of $\mathbf{v}$, the stationary distribution the Jackson network $(A,\R^5)$, which is represented by the black dots. After specialization (right), we see that the network has similar behavior to the original network. Specifically, using the initial condition $\mathbf{\overline{x}}_0 = [10,20,30,40,50,0]$, the dynamics on the network converge to $\mathbf{\overline{v}}$, the stationary distribution of $\left(\overline{A},\R^6\right)$, and the load at node 5 decreases.}
      \label{fig:Dynamics_Example}
     \end{subfigure}
\end{figure}

\begin{example}[\textbf{Minimal Specialization}] \label{Ex_Min_Spec}
Consider the graph $G = (V,E,W)$ in Figure \ref{fig:SpecEx} (left) with vertex set $V = \{1,2,3,4,5\}$. When specialized over the edge $e_{54} \in E$, the result is the graph $\overline{G} = (\overline{V},\overline{E},\overline{W})$ shown in Figure \ref{fig:SpecEx} (right) with vertex set $\overline{V} = \{1,2,3,4,5,5.1\}$, where $\overline{i} = 5.1$ is the copy of vertex $i = 5$. The dynamics of the original and specialized Jackson networks $(A, \R^5)$ and $(\overline{A},\R^6)$ are shown in Figure \ref{fig:Dynamics_Example}, left and right, respectively. 

As $A \in [0,1]^{5 \times 5}$ is primitive, the Jackson network $(A,\R^5)$ has the stationary distribution \[\mathbf{v} = [.0337838,.101351,.314189,.212838,.337838]^\intercal.\] Similarly, $\overline{A} \in \R^{6 \time 6}$ is primitive and the associated Jackson network has the stationary distribution \[\overline{\mathbf{v}} = [.0337838,.101351,.314189,.212838,.135135,.202703]^\intercal.\]
In Figure \ref{fig:Dynamics_Example} the initial conditions $\mathbf{x}^{(0)} = [10,20,30,40,50]^\intercal$ and $\overline{\mathbf{x}}^{(0)} =[10,20,30,40,50,0]^\intercal$ of the systems $\left(A,\R^5\right)$ and $\left(\overline{A},\R^6\right)$, respectively, lead to similar dynamics. On the left, $A^{k} \mathbf{x}_0$ is calculated for $k = 0, 1,\dots, 100$, where $\mathbf{x}_0 = [10,20,30,40,50]$. We see that the long-term dynamics approach $\|\mathbf{x}_0\|_1 \mathbf{v}$. After minimal specialization, we calculate $\overline{A}^{k} \mathbf{\overline{x}}_0$ for $k = 0, 1,\dots, 100$, where $\mathbf{\overline{x}}_0 = [10,20,30,40,50,0]$. Again, the asymptotic state of the system is $\|\mathbf{\overline{x}}_0\|_1 \overline{\mathbf{v}}$. Notice that $\|\mathbf{v}\| = \|\overline{\mathbf{v}}\|$, which is not a coincidence. This follows from Lemma \ref{Eigenvector}, and is explored further in Section \ref{interplay}.
\end{example}

The second deviation we propose from the original specialization models is how we choose where to evolve the graph structure of the Jackson network.. Previously, the structure was evolved stochastically by choosing vertices randomly from the graph to specialize \cite{Specialization, Spectral, Synchronization}. Here our goal is to use the dynamics on the network to determine where to specialize the structure.



\section{Coevolution of Structure and Dynamics}\label{interplay}
In real-world systems, the changing topology of the network is driven by the need to optimize the network's function, which is often related to moving quantities efficiently through the network. Dynamical processes such as traffic flow, information transfer, etc., put pressure on the network's topology to evolve in specific ways. In order to model this behavior, we use the dynamics on the network, or the dynamical processes the network models, to determine where to modify the network topology.

The node that experiences the highest traffic volume in a network is a natural candidate for the part of the network under the most stress. The idea is that once specialized, this node and its specialized copy now share the load, lessening the stress on the original node. A complicating factor in most real-world networks is that it may be difficult to determine which node experiences the most traffic. This may be the case, for instance, if the network dynamics are irregular, e.g., aperiodic, chaotic, etc. The reason we choose Jackson networks for our initial coevolution model is that, under mild assumptions, there is a clear hierarchy of which nodes experience more traffic. This is given by the network's stationary distribution.

\subsection{Integrated Specialization Model} \label{evec}

Suppose $(A,\R^n)$ is a Jackson network where $A \in [0,1]^{n \times n}$ is primitive. Then $A$ has a unique stationary distribution $\mathbf{x} = [x_1, x_2, \dots, x_n]^T \in \R^n$, where $x_i$ is the \emph{asymptotic use} of $i \in V$ in the associated network or graph $G = (V,E,W)$. The node that experiences the \emph{maximal asymptotic load} is the node $i$ such that $x_i = \max_\ell\{x_\ell\}_{\ell=1}^n$. To specialize the network relative to its dynamics, we choose node $i$ with maximal asymptotic load that has more than 2 outgoing edges. Given $i \in V$ we choose a node $j \in V, j \neq i$, such that $W(e_{ij}) = \max\{W(e_{i\ell}) | e_{i\ell} \in E, \ell \neq i\}$, i.e., we choose the edge that transitions the most traffic away from node $i$. Specializing $(A,\R^n)$ over node $i$ with edge $e_{ij}$ results in the specialized Jackson network $(\overline{A},\R^{n+1})$, which we refer to as the \emph{minimal dynamic specialization} of $(A,\R^n)$. 

We note that if there is a tie for the node with maximal asymptotic load, say nodes  $i,k \in V$, specializing $i$ with edge $e_{ij} \in E$ and then specializing $k$ with edge $e_{kh} \in \overline{E}$ will result in a different (non-isomorphic) graph structure than if the network is specialized in the opposite order. However, the asymptotic dynamics for all nodes after these two minimal dynamic specializations, in either order, will be the same. Similarly, if there is a tie for the highest edge weight, the graph structure will be different depending on which edge is chosen, but the resulting asymptotic dynamics for all nodes will be the same. In practice, a tie rarely occurs. If it does, we randomly choose one of the nodes (or edges) that is part of the tie to be used in the minimal dynamics specialization process.

If $A \in [0,1]^{n \times n}$ is primitive, by Theorem \ref{stable} it is possible to sequentially specialize the primitive Jackson network $\left(A,\R^n\right)$ via minimal dynamic specialization. For such networks we can define the following coevolving Jackson model which integrates both the dynamics on and the dynamics of the network.
\begin{definition}[\textbf{Integrated Specialization Model}]
    Let $(A_0,\R^n)$ be a Jackson network where $A_0 \in [0,1]^{n \times n}$ is primitive. We define the sequence of Jackson networks $\left\{\left(A_m, \R^{n+m}\right)\right\}_{m=0}^\infty$ to be the \emph{integrated specialization model} with initial Jackson network $(A_0,\R^n)$ and $A_{m+1} = \overline{A}_m$ for $m \geq 0$ is the minimal dynamic specialization of $A_m$.
\end{definition}

\subsection{Irreducibility}

Having a Jackson network $(A,\R^n)$ where $A \in [0,1]^{n \times n}$ is primitive is a strong condition, and a property that is difficult to establish since it typically involves calculating eigenvalues, taking large powers of matrices, calculating path lengths, etc., all of which are computationally intensive for large networks. However, a condition that is  more reasonable is for $A$ to be irreducible, which is equivalent to having a strongly-connected graph (see, for instance, \cite{horn2012matrix}). 

Most real-world networks have a largest strongly-connected component that comprises the majority of the network \cite{newman2018networks}. Thus, the underlying graph structure of the network is strongly connected if we restrict our attention to the graph's largest strongly-connected component. This guarantees that the network's adjacency matrix $A$ is irreducible. With irreducibly, we still have $\rho(A)$ as an algebraically simple eigenvalue with positive leading eigenvector, but we lose having a unique stationary distribution (i.e. stable asymptotic dynamics). This positive leading eigenvector is the network's \emph{eigenvector centrality}, which gives a ranking of the nodes that takes into account the importance of a node relative to the importance of its neighbors \cite{newman2018networks}. A necessary condition for primitivity is irreducibly, so in the setting of primitivity, we still have a notion of \emph{eigenvector centrality}. In fact, if we have a primitive, stochastic matrix $A$, the eigenvector centrality is equivalent (up to scaling) to the associated Jackson network's stationary distribution. 

Similar to our minimal dynamic specialization, we can use this ranking to determine which node to use in minimal dynamic specialization, i.e., we choose a node with maximal eigenvector centrality, eligible for minimal dynamic specialization, and choose the edge in the same manner as in minimal dynamic specialization. Moreover, Lemma \ref{strong} tells us that if $A$ is strongly connected, then so is $\overline{A}$. Thus, like with the integrated specialization model, we use can minimal dynamic specialization to create a sequence of irreducible Jackson networks. 

\begin{example}[\textbf{Minimal Dynamic Specialization}]
Consider the Jackson network $(A,\R^5)$ shown in Figure \ref{fig:SpecEx} (left). The Jackson network's stationary distribution is
$$
    \mathbf{v} = \begin{bmatrix}
        .0337838,
        .101351,
        .314189,
        .212838,
        .337838
    \end{bmatrix}^\intercal,
$$
so $\max\{x_i\} = x_5$ is the maximum load and $\max\{W(e_{5\ell})\} = W(e_{54})$. Thus the minimal  dynamic specialization of $(A,\R^5)$ is the Jackson network $(\overline{A},\R^6)$ shown in Figure \ref{fig:SpecEx} (right). Note that the specialized network has the stationary distribution 
$$
    \overline{\mathbf{v}} = \begin{bmatrix}
        .0337838,
        .101351,
        .314189,
        .212838,
        .135135,
        .202703
    \end{bmatrix}^\intercal.
$$

Figure \ref{fig:Dynamics_Example} shows the dynamics on the original network from Figure \ref{fig:SpecEx} (left) and the specialized network (right). Over time, the quantities at each node tend to converge to the (scaled) quantities of $\mathbf{v}$, represented as black dots. After specialization, the dynamics on the network behave in the same way, converging to (a scaled) $\mathbf{\overline{v}}$. The the total traffic load in these systems is conserved, but the 
maximal load of $\left(\overline{A},\R^6\right)$ is reduced.
\end{example}

Lemma \ref{Eigenvector} gives us a way of tracking how the maximum eigenvector centrality changes with the integrated specialization model. In particular, the sequence $\left\{\left(A_m,\R^{m+n}\right)\right\}_{m=0}^\infty$ has the sequence of leading eigenvectors $\mathbf{v}^{(0)},\mathbf{v}^{(1)},\mathbf{v}^{(2)} \cdots $, where $$v^{(k+1)}_{\ell} = \begin{cases}
     (1 - w) v^{(k)}_{i}  & \text{  if } \ell = i \\
     wv^{(k)}_{i}  & \text{  if } \ell = \overline{i} \\
     v^{(k)}_{\ell}  & \text{  else } \\
\end{cases}$$
with $\|\mathbf{v}^{(0)}\|_1 = \|\mathbf{v}^{(1)}\| = \dots \|\mathbf{v}^{(k)}\| = \dots  $. This allows us to recursively compare the maximum eigenvector centrality as we specialize. Consequently, Lemma \ref{Eigenvector} describes the evolution of the eigenvector centrality vector as a graph $G$ with a column stochastic adjacency matrix is specialized under minimal dynamic specialization. Specifically, we have $$\max_{1\leq k \leq n} v_k^{(m)} \geq \max_{1\leq k \leq n+1} v_k^{(m+1)} \ \ \text{ for }  m \geq 0$$
In our sequence of leading eigenvectors. Thus, we are targeting and decreasing the areas of high stress or high importance on the network.

\begin{figure}[h!]
    \centering
    \includegraphics[width=1\textwidth]{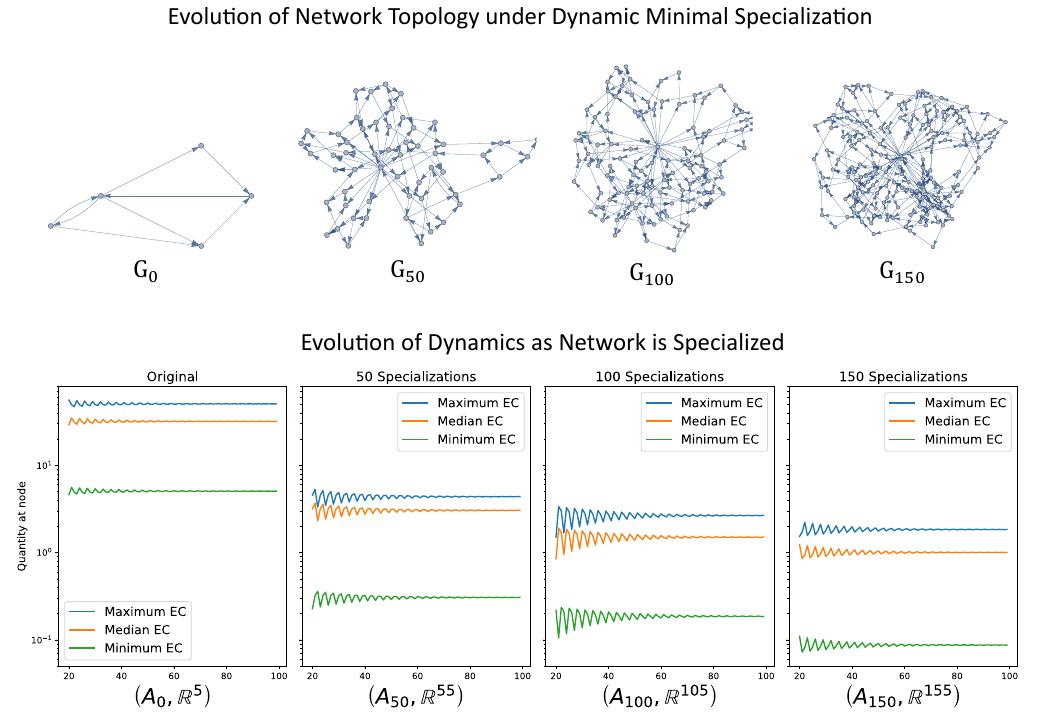}
    \caption{The long-term dynamics associated with the Jackson network in figure \ref{fig:SpecEx} is shown in part, where the dynamics of the network are shown above and the dynamics on the network are shown below. The dynamics are generated using the integrated specialization model. Each plot on the bottom has the dynamics for the node with the highest eigenvector centrality, the node with the median eigenvector centrality, and the node with minimum eigenvector centrality, plotted on a log scale. As the network is repeatedly specialized, these quantities decrease by an order of magnitude.}
      \label{fig:evolve_Dynamics}
\end{figure}

\begin{example}
    Sequentially specializing the Jackson network $\left(A,\R^5\right)$ in Example \ref{fig:SpecEx} using the integrated specialization model results in the sequence $\left\{\left(A_m, \R^{5+m}\right)\right\}_{m=0}^\infty$. Figure \ref{fig:evolve_Dynamics} (bottom) shows the long-term dynamics for $m = 0,50,100,150$. Each plot has the dynamics for the node with the highest eigenvector centrality, the node with the median eigenvector centrality, and the node with minimum eigenvector centrality, plotted on a log scale. As the network is specialized, these quantities are getting increasingly closer together and decrease by at least an order of magnitude. That is, as we specialize high-stress areas of the network, we are decreasing network's maximal asymptotic load, and repeatedly doing so creates more equidistributed asymptotic loads.
\end{example}

\section{Real-World Properties of the Integrated Specialization Model}\label{properties}
There are numerous models that generate networks which achieve specific real-world properties. These properties include (i) right-skewed degree distributions, (ii) sparsity, (iii) the small-world property etc. However, many models only exhibit one or two of these properties. For example, an Erd\"{o}s-R\'{e}nyi network has a giant component, a Barab\'{a}si-Albert network has a right-skewed degree distribution, and a Watts-Strogatz network exhibits the small-world property \cite{newman2018networks,durrett2007random}. In this section, we give statistical evidence that the integrated specialization model exhibits each of these real-world properties. In our experiments, we begin with a Jackson network $(A_0, \R^{25})$ given by an Erd\"{o}s-R\'{e}nyi graph $G = (V,E,W)$, with $|V| = 25$ nodes and a density of $G_{den} = .25$, where edge weights are uniformly assigned and normalized so outgoing edges sum to 1. We use the integrated specialization model to produce the sequence $\left\{\left(A_m, \R^{25 + m}\right)\right\}_{m=0}^{500}$, growing the network to $525$ nodes. For this sequence of graphs, we collect statistics (i)-(iii) and repeat this experiment for 100 such initial Jackson networks. The averaged statistics with standard-deviations are described in the following subsections.

\subsection{Degree Distribution}
Real-world networks exhibit a right-skewed degree distribution, meaning many nodes have low degree and few nodes have high degree \cite{newman2018networks}. Since we are analyzing directed networks in our numerical experiment, we consider both the in-degree and out-degree distributions.

\begin{figure}[h!]
    \centering
    \includegraphics[width=.8\textwidth]{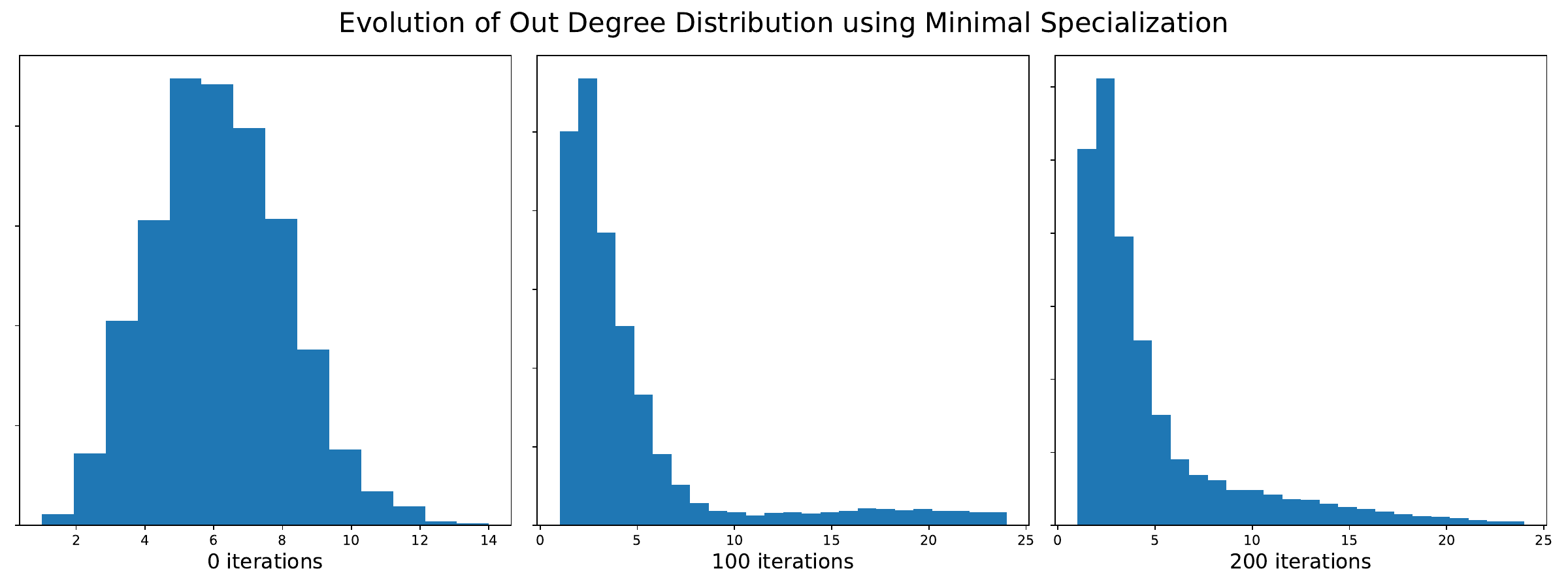}
    \caption{Averaged results describing how the out-degree distribution of the network evolves as repeated minimal dynamic specialization is performed. As the network evolves, the out-degree distribution becomes increasingly right-skewed, a hallmark of real-world networks.}
      \label{fig:Degree}
\end{figure}

Figure \ref{fig:Degree} shows our results regarding how the out-degree evolves, on average, over our 100 trials. The left histogram,which has a distinct binomial shape, is the average histogram for 100 directed Erd\"{o}s-R\'{e}nyi graphs. This binomial shape is very unlike the right-skewed distribution found in real-world networks \cite{newman2018networks}. The middle panel shows the averaged histogram for the networks $(A_{100},\R^{125})$ over 100 simulates, and the right panel is the averaged histogram of $(A_{200},\R^{225})$ over 100 simulations. As these graphs are specialized, we see, on average, an increasingly right-skewed degree distribution, i.e., an increasingly more real-world like degree distribution.

There is a simple heuristic that helps explain why we see a right-skewed out-degree distribution. When we use the integrated specialization model, a node is created with only one outgoing edge. As we sequentially specialize, it's possible for nodes with a small number of out-going edges to gain more outgoing edges, but we are still adding a node with an out-degree of 1 at each iteration.

 We note that one drawback of the integrated specialization model is that the in-degree distribution does not evolve into a right-skewed distribution. 
 To solve this, we could alternate between specializing $G$ and $G^T$, where $G^T$ is the graph with adjacency matrix $A^T$. However, we would not have the same theoretical properties as the integrated specialization model defined here.    
\begin{figure}
     \centering
     \begin{subfigure}[t]{0.32\textwidth}
        \vskip 0pt
         \includegraphics[width=\textwidth]{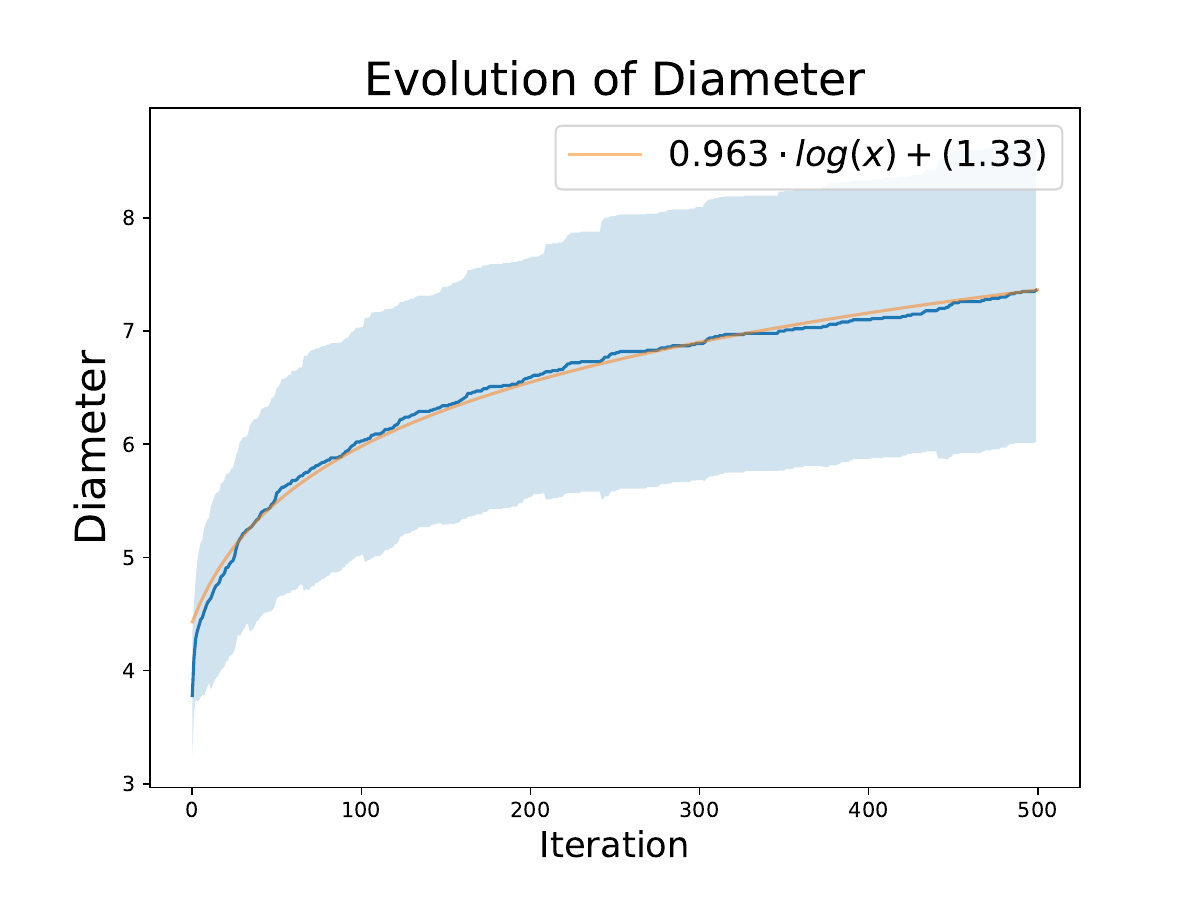}
         \caption{Averaged results describing how the diameter of the network evolves using the integrated specialization model. One standard deviation is shown with the shaded region. The orange curve is a logarithmic equation fitted to the data. The trend of the diameter is logarithmic, i.e., the model exhibits the \textit{small-world property}, a phenomenon that is typical of real-world networks.}
         \label{fig:smallworld}
     \end{subfigure}
     \hspace{.1pt}
     \begin{subfigure}[t]{0.32\textwidth}
        \vskip 0pt
         \includegraphics[width=\textwidth]{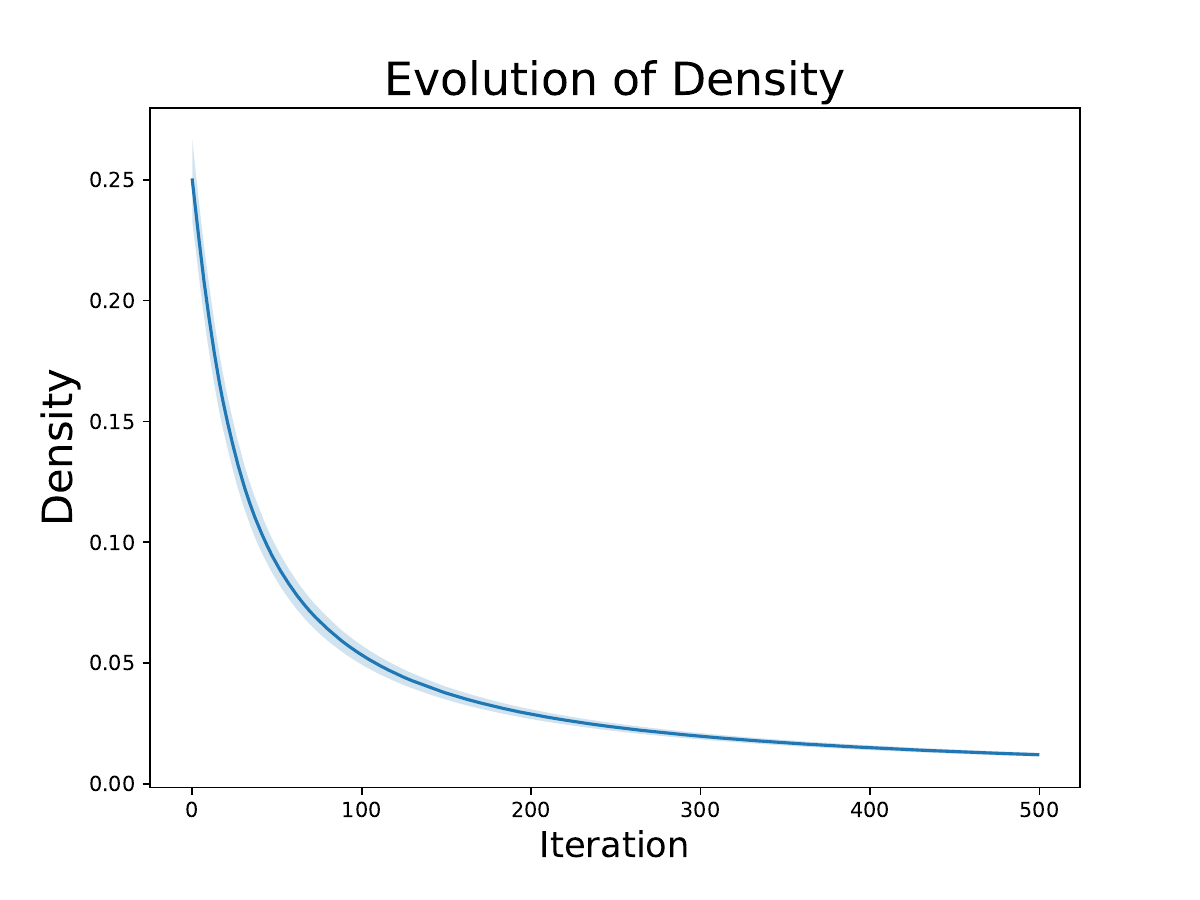}
         \caption{Averaged results describing how the density of the network evolves using the integrated specialization model. One standard deviation is shown, with the maximum standard deviation as 0.017 and the minimum as  0.0008. As the network evolves, the density decreases overall, a hallmark of real-world networks.}
         \label{fig:Density}
     \end{subfigure}
     \hspace{.1pt}
     \begin{subfigure}[t]{0.32\textwidth}
        \vskip 0pt
         \includegraphics[width=\textwidth]{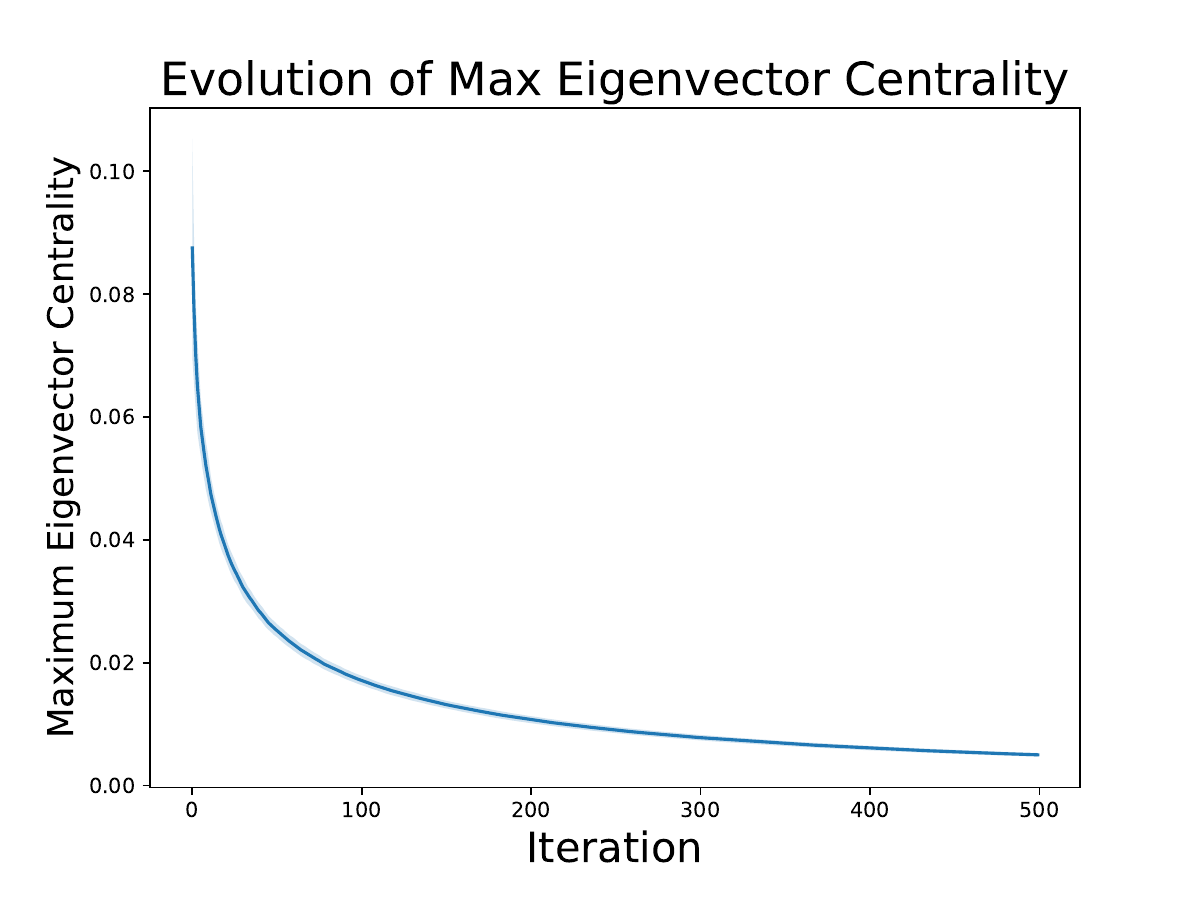}
         \caption{Averaged results describing how the maximum eigenvector centrality of the network evolves using the integrated specialization model. One standard deviation is shown, with the maximum standard deviation as 0.0178 and the minimum as  0.0003. As the network evolves, the maximum eigenvector centrality decreases, indicating that we are reducing of high stress in the network.}
         \label{fig:Evec}
     \end{subfigure}
\end{figure}

\subsection{Small-World Property} \label{small-world}
The diameter of a network is the network's largest geodesic, i.e., its longest shortest path. Intuitively, it's the furthest traffic, information, etc., will need to travel in a network.
It has been observed that as a real-world network evolves over time, its diameter grows logarithmically. This phenomenon is known as the \textit{Small-World Property} \cite{newman2018networks}.

 Figure \ref{fig:smallworld} shows how the diameter of the sequence $\left\{\left(A_m,\R^{25 + m}\right)\right\}_{m=0}^{500}$ grows, averaged over 100 such sequences. The blue curve is the evolution of the average diameter, with the shaded region representing one standard deviation. The orange curve is fitted to the blue curve. The growth appears to be logarithmic, suggesting that the integrated specialization model has the Small-World Property, at least for initial networks with an Erd\"{o}s-R\'{e}nyi topology.

\subsection{Density}
The density of a directed network $G = (V,E,W)$ is defined to be 
\[G_{den} = \frac{m}{n(n-1)},\]
where $m = |E|$ is the number of edges and $n = |V|$. The density can be thought of as the ratio of edges to possible edges in a graph. If the ratio tends to $0$ as the network grows, the network is said to be \emph{sparse}; otherwise, it is said to be \emph{dense}. A hallmark of real-world networks is that they appear to be sparse when compared to random networks \cite{Real-world}. 

Figure \ref{fig:Density} shows how the density changes as a network grows using the integrated specialization model. For the graphs we consider, the average density begins at $G_{den} = .25$ and the sparsity rapidly drops toward zero, at a nearly exponential rate. One standard deviation is shaded, but is too small to observe, indicating a very constrained evolution towards sparsity.

\subsection{Maximum Eigenvector Centrality}\label{max_evec}

In subsection \ref{evec}, we described the role eigenvector centrality has in our dynamical model. To reiterate, real-world networks tend to specialize in areas the network with high traffic or stress. The integrated specialization model is designed to specialize nodes with high eigenvector centrality to maximize dynamic functionality by reducing high-stress areas.
A consequence of Lemma \ref{Eigenvector} is that the maximal eigenvector centrality, i.e., the maximal load, of a network cannot increase as the network is specialized.

Figure \ref{fig:Evec} shows the averaged results for how the maximal asymptotic load evolves with the integrated specialization model. One standard deviation is shown, but is very small. On average, we see a rapid decrease in the maximum eigenvector centrality; thus, statistically, we do much better than the theory informs, efficiently targeting areas of high traffic in the network and successfully reducing the network's maximal load, i.e. areas of stress.

\section{Comparison to Other Models}\label{comapre_models}

The most novel feature of the integrated specialization model is that it uses the dynamics on the network to determine where to evolve the structure of the network. In the previous section, we presented statistical evidence that the integrated specialization model creates real-world structural properties. To understand how this coevolution leads to structural differences, we compare the integrated specialization model to two other models: (i) \emph{random minimal specialization} and (ii) the \emph{\textit{Barab\a'si-Albert}} (BA) model.

In random minimal specialization, we remove the integrated specialization model's dependence on the network's dynamics by performing minimal specialization over a random node using its edge with the highest weight. This creates a sequence of Jackson networks $\left\{\left(R_m, \mathbb{R}^{m + n}\right)\right\}_{m \geq 0}$. In the Barab\a'si-Albert model we consider a single node preferentially added via a single edge at each step, resulting in a sequence of simple graphs $\{B_m\}_{m \geq 0}$, where $|B_m| = m + n$.

\begin{figure}
     \centering
     \begin{subfigure}[t]{0.45\textwidth}
        \vskip 0pt
         \includegraphics[width=\textwidth]{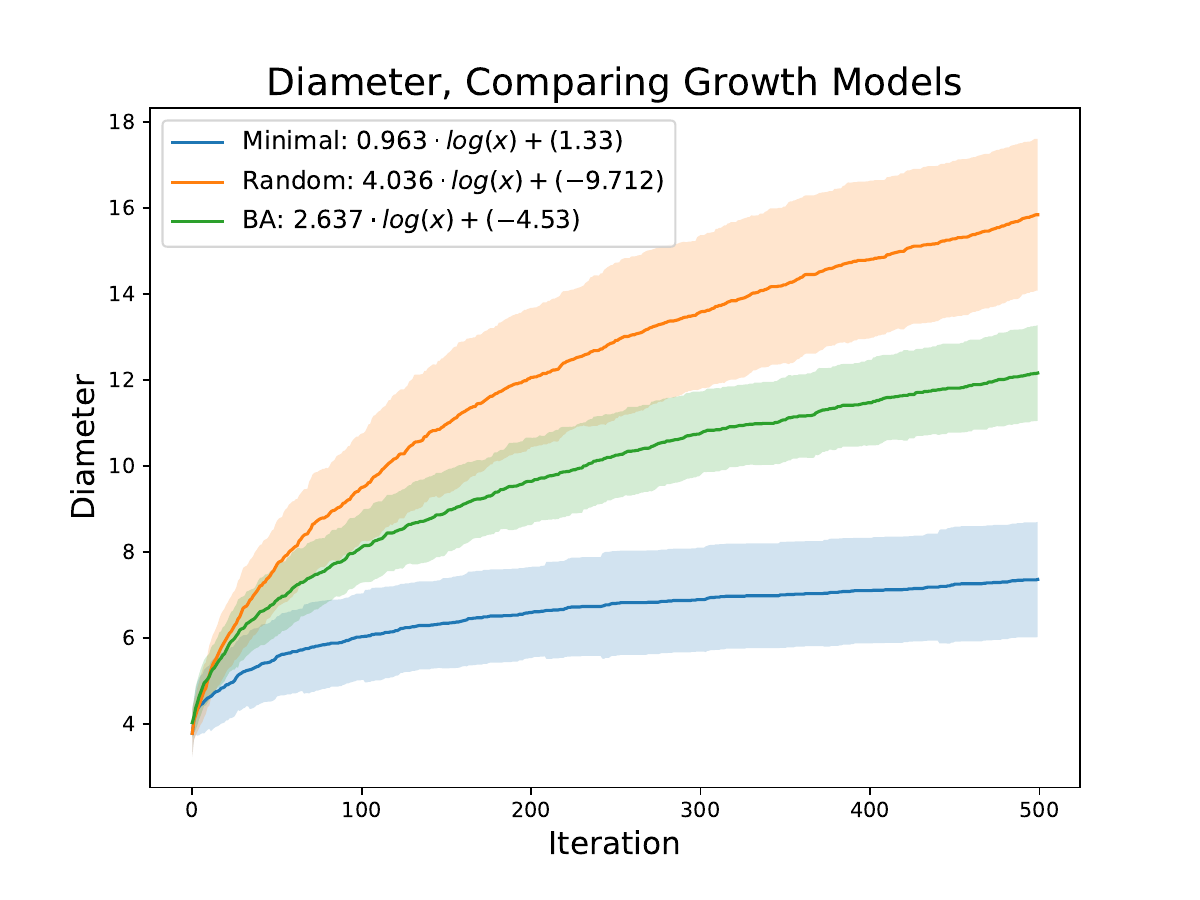}
         \caption{Figure comparing the evolution of the diameter for the three growth models, integrated specialization, random minimal specialization, and Barab\a'si-Albert model (adding one edge at a time). 100 simulations of networks grown to 525 nodes were averaged. A line of best fit is shown for each model. In each case the models appear to have logarithmic growth, i.e., the \emph{small-world} property.}
         \label{fig:Compare_Diameter}
     \end{subfigure}
     \hspace{.1pt}
     \begin{subfigure}[t]{0.45\textwidth}
        \vskip 0pt
         \includegraphics[width=\textwidth]{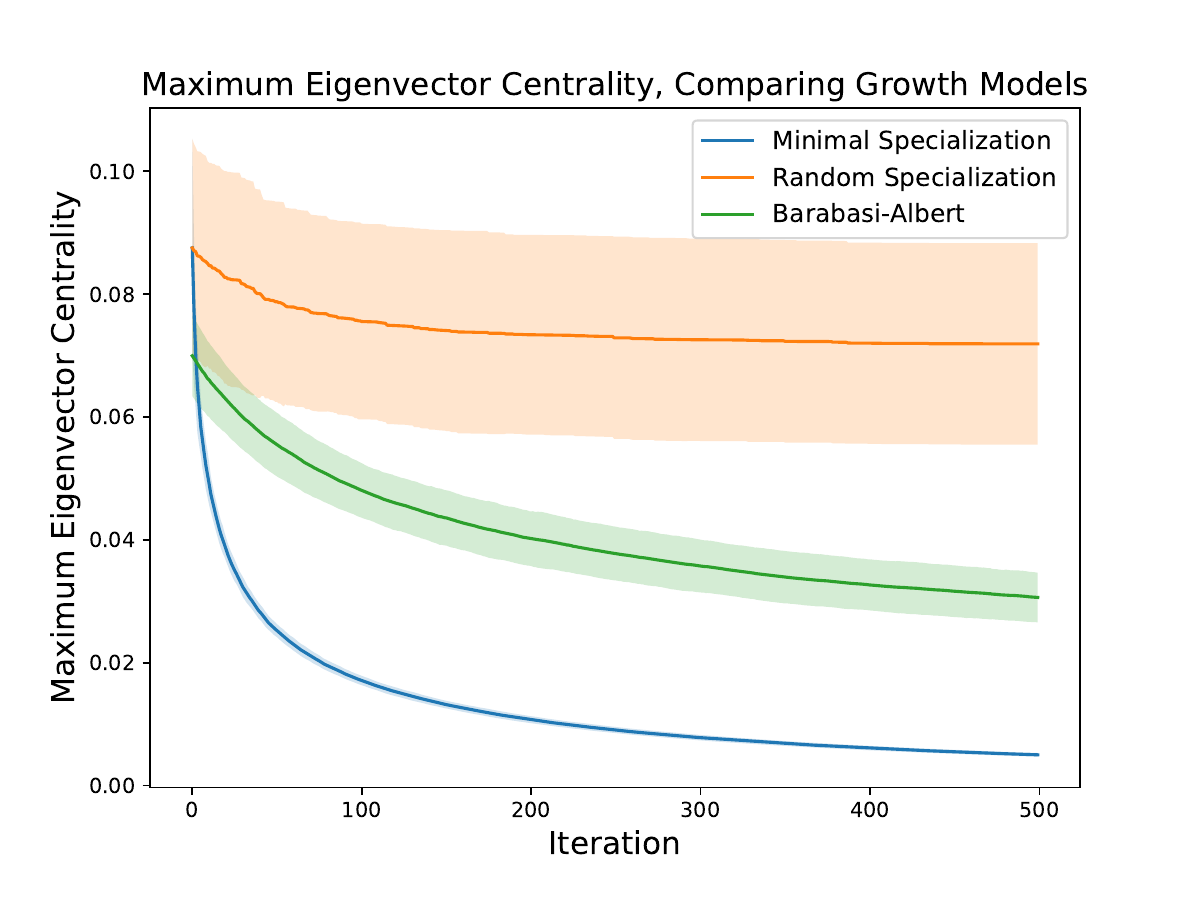}
         \caption{Figure comparing the evolution of the maximum eigenvector centrality for the three growth models, integrated specialization, random minimal specialization, and Barab\a'si-Albert model (adding one edge at a time). 100 simulations of networks grown to 525 nodes were averaged. For the maximum eigenvector centrality over time, the integrated specialization model does a better job of decreasing the maximum eigenvector centrality and does so with little variance between simulations. Thus, the integrated specialization model is more efficient at targeting high-stress areas of the network and relieving that stress.}
         \label{fig:Compare_Evec}
     \end{subfigure}
\end{figure}

As in the previous section, we begin with an Erd\"{o}s-R\'{e}nyi graph $G_0 = (V,E,W)$ with $|V| = 25$ and $G_{den} = .25$ for each of the integrated specialization, random minimal specialization, and Barab\a'si-Albert Models (directed for the first two and undirected for BA). For each model, we evolve the networks 500 times to create the sequences $\left\{\left(A_m,\R^{n+m}\right)\right\}_{m = 0}^{500}$, $\left\{\left(R_m,\R^{n+m}\right)\right\}_{m = 0}^{500}$, and $\left\{G_m\right\}_{m = 0}^{500}$, respectively. This experiment is performed 100 times for each model and the data is averaged. We note that for each of these models, the evolution of the density is essentially the same, rapidly decreasing to zero at nearly the same rate. Similarly, the degree distributions are fundamentally the same, evolving to a right-skewed degree distribution. However, the average growth of the diameter of the three models exhibit the Small-World property but show different growth rates (see Figure \ref{fig:Compare_Diameter}).

In the networks we consider, the maximum eigenvector centrality corresponds to the node that, on average, has the most information or traffic, i.e. the maximal load. Thus, a decrease in maximum eigenvector centrality corresponds to a decrease in areas of high stress on the network. Figure \ref{fig:Compare_Evec} shows the evolution of the maximum eigenvector centrality, or maximal load, in each of the three growth models, where each eigenvector is normalized by the 1-norm. The integrated specialization model rapidly decreases the maximum load, which is expected. Random minimal specialization and BA do not decrease as rapidly, or reach as low of a value as the integrated specialization model. Moreover, the variance across simulations for the integrated specialization model is negligible, while there is higher variance in the other two models. 

The differences in the evolution of the diameter and maximal load give evidence that the integrated specialization model creates efficient networks, both in terms structure and dynamics. Specifically, the integrated specialization model creates networks with small diameter, so that distances across the network are minimized, and equidistributed traffic, so that traffic bottlenecks are reduced throughout the network. Moreover, the negligible variance between trials for the integrated specialization model suggests that we achieve these results in a near-optimal way.

\section{Equitable Partitions} \label{ep ch}
A hallmark of real-world networks is the high occurrence of symmetric structures \cite{MACARTHUR20083525}. Each such symmetry is given by an equitable partition, which is a generalization of the notion of a graph symmetry. Historically, equitable partitions are defined for \emph{simple graphs}: unweighted, undirected graphs without loops. However, equitable partitions can be defined for unweighted directed graphs $G = (V,E)$ as follows \cite{Synchronization}:

\begin{definition}[\textbf{Equitable Partition}]
Let $G = (V,E)$ be a graph with adjacency matrix $A = A(G)$. Let $\pi = \{V_1,V_2, \cdots V_k\}$ be a partition of the vertices $V$. Then $\pi$ is an \emph{equitable partition} if the sum 
\begin{equation}\label{ep}
    \sum_{j \in V_b} A_{ij} = D_{ab} 
\end{equation}
is constant for any $i \in V_a$.
If $|V_i| = 1$, we call $V_i$ trivial, else we call it non-trivial. We call $\pi$ trivial if each element in $\pi$ is trivial. We call the matrix $D \in \N^{k \times k}$ the divisor matrix of A associated with $\pi$, and $G_{\pi}$ the divisor graph associated with $D$. 
\end{definition}\label{ep_def}

We emphasize that our definition uses the unweighted adjacency matrix, meaning we are focusing on the topology of the network and not the edge weights. A Jackson network $(A,\R^n)$ has an equitable partition $\pi$ if its associated unweighted graph $G = G(A)$ where $G = (V,E)$ has the equitable partition $\pi$.

\begin{figure}[h!]
    \centering
    \includegraphics[width=1\textwidth]{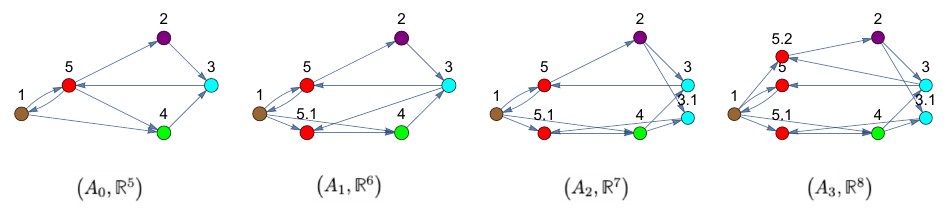}
    \caption{An example illustrating the formation of non-trivial equitable partitions. On the right is the original network. From left to right, node 5 is specialized, with node 5.1 as its copy. These are in the same equitable partition element (red). Next, node 3 is specialized with node 3.1 as its copy, which are in the same partition element (cyan). Finally, node 5 is specialized again, with node 5.2 as its copy, increasing the partition element size from two to three (red).}
      \label{fig:EP_Evolve}
\end{figure}

The main result of this section is that minimal specialization creates and preserves nontrivial equitable partition elements. Later, we study the consequences of repeated minimal specialization on the size and type of equitable partitions a specialized network has (see Corollary \ref{size} and Figure \ref{fig:EP_Evolve}). 

\begin{theorem}[\textbf{Preservation of Equitable Partitions}] \label{Pres}
Let $\left(\overline{A}, \R^{n+1}\right)$ be the minimal specialization of the Jackson network $(A,\R^n)$ over any eligible vertex $i$ with edge $e_{ij}, i \neq j$. If $(A,\R^n)$ has an equitable partition $\pi = \{V_1,V_2, \cdots V_k\}$, then $\left(\overline{A},\R^{n+1}\right)$ has an equitable partition $\overline{\pi}  = \{\overline{V}_1,\overline{V}_2, \cdots \overline{V}_k\}$ where
$$
\overline{V}_a = \begin{cases}
    V_a \cup \{\overline{i}\} \ \ \ &\text{ if } i \in V_a \\
    V_a \ \ \ &\text{ otherwise}
\end{cases}
\; \; \; \; \; \; \text{ for } a = 1,2, \cdots, k
$$
where $\overline{i}$ is the node created during minimal specialization.
\begin{proof}
We first note that Definition \ref{ep_def} is equivalent to saying we can partition our adjacency matrix so that the row sum in each partition is constant. Thus, we will consider the partitioned adjacency matrices corresponding to $\pi$ and $\overline{\pi}$. Without loss of generality, let $ j \in V_{k-1}$ and $i \in V_k$. The partitioned adjacency matrix is

\begin{align}
A = \left[\begin{array}{c| c|c| c }
A_{11}  & A_{12} & \cdots&  A_{1k}  \\
\hline
\vdots & \vdots  &\vdots  &\vdots \\
\hline
A_{(k-1)1}  & \cdots  & A_{(k-1)(k-1)} & A_{(k-1)k} \\
\hline
A_{k1}  & A_{k2}  & A_{k(k-1)}  & A_{kk}
\end{array}\right] \in \R^{n \times n}
\end{align}

Where each $A_{ii}$ is a block matrix that represents the connections within a part and each $A_{ij}, i\neq j$ a block matrix that represents the connections from part $i$ to part $j$. 

 Let $\tilde{A}_{(k-1)k}$ denote the matrix that has the same entries as $A_{(k-1)k}$ except the entry corresponding to $e_{ij}$ in the adjacency matrix is changed from a 1 to a 0. This represents deleting the edge from $i$ to $j$, which is done during minimal specialization. 

 Let $\tilde{A}_{k\ell}$ for $1 \leq \ell \leq k$ denote the matrix that has the same entries as $A_{k\ell}$ but now we add an extra row that is a copy of the row corresponding to node $i$. This represents $\overline{i}$ having the same in-edges as $i$.

Since $\overline{i}$ has only one outgoing edge, which is to node $j$, the column in $\overline{A}$ corresponding to $\overline{i}$ is all zeros except a 1 in the $jth$ entry.

Thus, the partitioned adjacency matrix of $\overline{A}$ is

\begin{align}
\overline{A} = \left[\begin{array}{c| c|c| c c }
A_{11}  & A_{12} & \cdots&  A_{1k} & \Medzero \\
\hline
\vdots & \vdots  &\vdots  & &\vdots \\
\hline
A_{(k-1)1}  & \cdots  & A_{(k-1)(k-1)} & \tilde{A}_{(k-1)k} & e_j \\
\hline
\tilde{A}_{k1}  & \tilde{A}_{k2}  & \tilde{A}_{k(k-1)}  & \tilde{A}_{kk} & \Medzero
\end{array}\right] \in \R^{(n+1) \times (n+1)}
\end{align}
where $e_j$ is the vector that is all zeros except for a 1 that corresponds to the row associated with $j$. In this form, it is clear that the upper left $(k-1) \times (k-1)$ block matrix is the as the upper left $(k-1) \times (k-1)$ block matrix of $A$, so the row sums are constant on each part. Moreover, in the last partition column, it is clear that partition rows $1$ to $k-2$ have constant row sum since we are only adding a column of zeros. For the $(k-1),k$ partition of $\overline{A}$, if we are not examining the row corresponding to node $j$, the row sum is the same as the row sum in $A_{(k-1)k}$. If we are looking at the row corresponding to $j$, then the row in $\tilde{A}_{(k-1)k}$ has the same entries as $A_{(k-1)k}$ except for the entry corresponding to $i$, which is $0$ for $\tilde{A}_{(k-1)k}$ and $1$ for $A_{(k-1)k}$. But, we have an additional $1$ in the row corresponding to $j$ in $\overline{A}$. Thus the row sums for $\begin{bmatrix}
    \tilde{A}_{(k-1)k} & e_j
\end{bmatrix}$ are the same as $A_{{k-1}k}$, and thus the row sums are constant. Finally, for the last partition row, since the rows of $\tilde{A}_{k\ell}$ are copies of the rows of $A_{k\ell}$ (some repeated), and we are only adding a column of zeros in the case of $\tilde{A}_{kk}$, the row sums are constant. 

Thus, the partitioned $\overline{A}$ has constant row sums on each part, so equivalently, $\overline{\pi}$ is an equitable partition.

We note that if $i$ and $j$ are in the same partition element, then the argument is similar, but instead of having an $\tilde{A}_{(k-1)k}$ and $\tilde{A}_{kk}$, there is just $\tilde{A}_{kk}$ with an extra row that is a copy of the row corresponding to $i$ in $A_{kk}$ and changing the appropriate entry from a $1$ to a $0$.
\end{proof} 
\end{theorem}

Using Theorem \ref{Pres}, it follows the the size of a network's equitable partition remains constant under minimal specialization, i.e., the following holds.

\begin{corollary}\label{size}
Let $\left\{\left(A_m,\R^{n + m}\right)\right\}_{m = 0}^\infty$ be a sequence of Jackson networks created via repeated minimal specialization. Denote the trivial partition of $(A_0,\R^n)$ by $\pi_0$. Then $(A_m,
R^{n+m})$ has an equitable partition $\pi_m$ where $|\pi_m| = |\pi_0|$ for all $m \geq 0$.

\begin{proof}
This follows from an inductive argument. Our base case is $\pi_0$ being the trivial equitable partition of $(A_0,\R^n)$. Thus $|\pi| = |V|$. Now assume inductively that for $(A_{m-1},\R^{m-1 + n})$, we have that $|\pi^{m-1}| = |V|$. Let $(A_{m},\R^{m + n})$ be the minimal specialization of $(A_{m-1},\R^{m-1 + n})$. From theorem \ref{Pres}, we have that $|\pi^{n}| = |\pi^{n-1}| = |V|$.
\end{proof}
\end{corollary}

The number of partition elements remains fixed as a network is specialized, meaning that as the network grows, the partition elements are what grow in size. A related question, is as the network grows, how do the sizes of the partition elements grow? For the sake of illustration, we consider this in the following example.

\begin{example} 
Figure \ref{fig:EP_Evolve} shows an example of an evolving Jackson network given by the sequence $\left\{\left(A_m, \R^{5 + m}\right)\right\}_{m=0}^3$. The Jackson networks are sequentially specialized over the vertices $i = 5,3, \text{ and } 5$ again, respectively. Here the original network  has the trivial equitable partition $\pi_0 = \{\{1\},\{2\}, \{3\}, \{4\}, \{5\} \}$. The result of specializing in this manner results in the partitions
\begin{align*}
    &\pi_1 = \{\{1\},\{2\}, \{3\}, \{4\}, \{5,5.1\} \} \\
    &\pi_2 = \{\{1\},\{2\}, \{3,3.1\}, \{4\}, \{5,5.1\} \} \\
    &\pi_3 = \{\{1\},\{2\}, \{3,3.1\}, \{4\}, \{5,5.1,5.2\}\},
\end{align*}
where each partition element is colored brown, purple, blue, green, and red, respectively. Note that $| \pi_0 | = | \pi_1 | = | \pi_2 | = | \pi_3 |$.

\end{example}
\subsection{Evolution of Equitable Partitions and Comparisons}

 In the previous section, we proved that minimal specialization either creates a new non-trivial partition element or grows the size of a non-trivial element. Thus, the percentage of non-trivial equitable partition elements never decreases. Here, we consider numerically the rate at which non-trivial elements are created. The experiments here are the same as those discussed earlier in Section \ref{comapre_models}, beginning with a directed Erd\"{o}s-R\'{e}nyi graph of $|V| = 25$ nodes, $G_{den} = .25$ density, and uniform, normalized edge weights for dynamic and random minimal specialization models. We also examined two  Barab\a'si-Albert models, one where we preferentially attach one node with one edge, and the other with one node and two edges. For these models, we begin with an undirected, unweighted Erd\"{o}s-R\'{e}nyi graph with $|V| = 25$ nodes, $G_{den} = .25$ density. We iterate each model 500 times, calculating the percentage of non-trivial equitable partition elements at each time step. We repeat this for 100 simulations and averaged the results.

 \begin{figure}
     \centering
     \begin{subfigure}[t]{0.45\textwidth}
        \vskip 0pt
         \includegraphics[width=\textwidth]{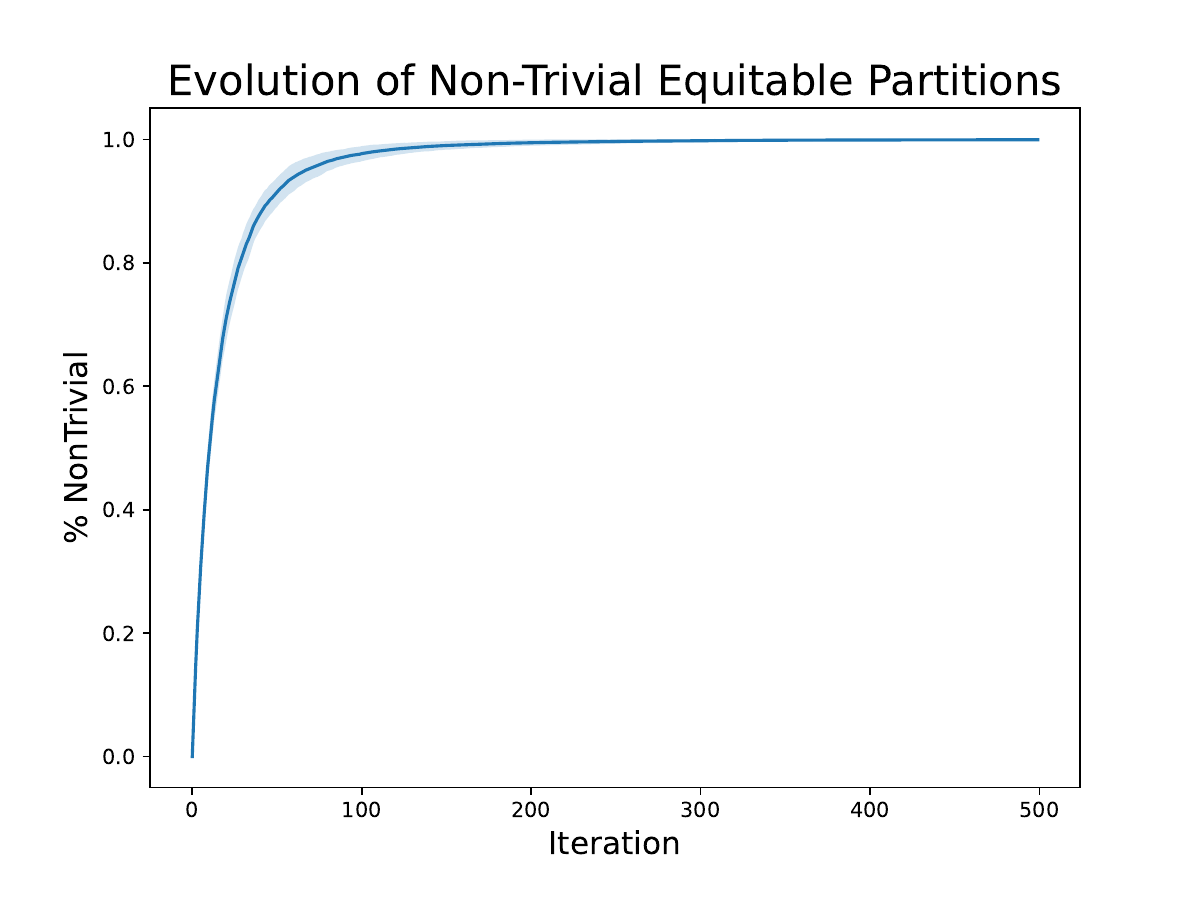}
         \caption{An example illustrating the evolution of the percentage of nontrivial equitable partition elements using the integrated specialization model. As the network evolves, the percentage of nontrivial equitable partition elements increases. Little standard deviation is observed, with the maximum being .0378 and the minimum being 0 (in every simulation, the final network had 100\% non-triviality). }
         \label{fig:EP}
     \end{subfigure}
     \hspace{.1pt}
     \begin{subfigure}[t]{0.45\textwidth}
        \vskip 0pt
         \includegraphics[width=\textwidth]{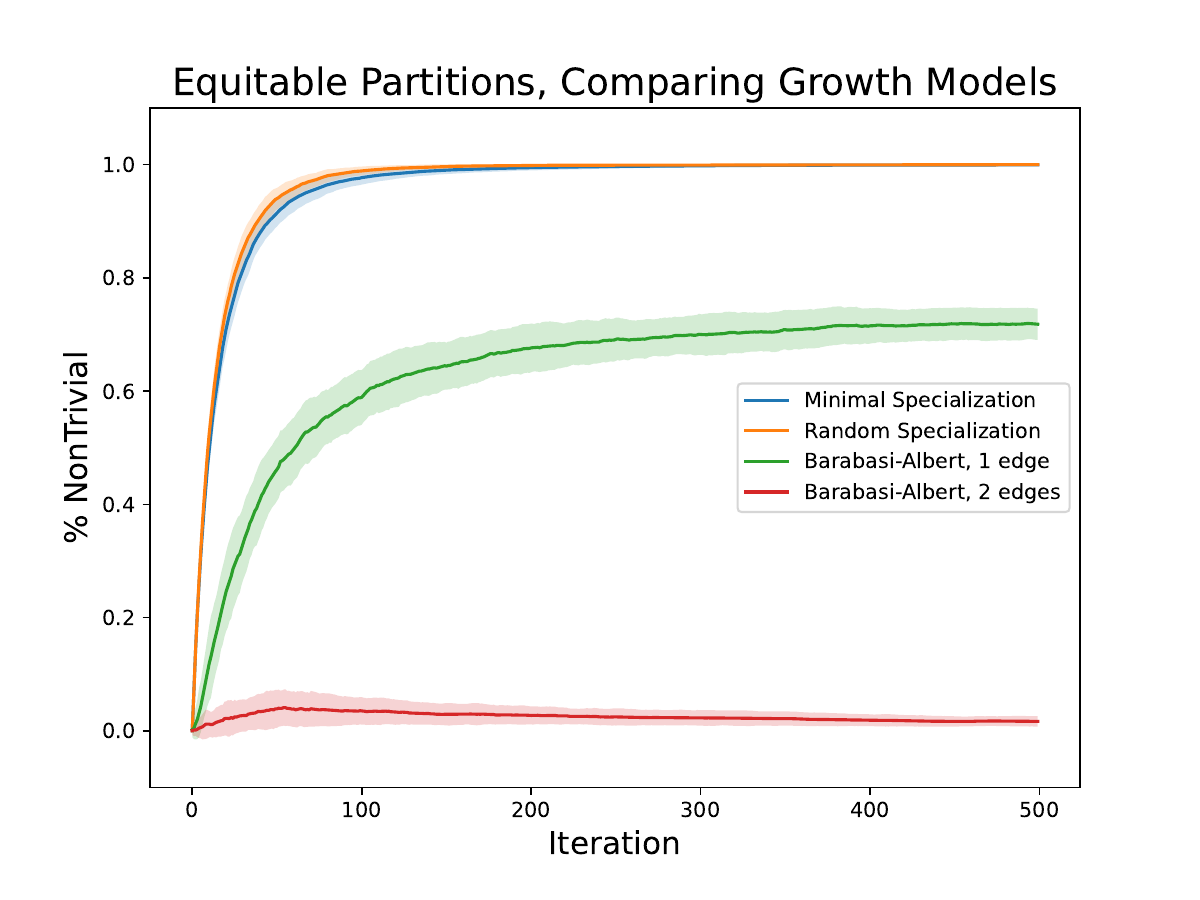}
         \caption{The evolution of the maximum percentage of non-trivial equitable partition elements for the growth models, the integrated specialization model, random minimal specialization, and Barab\a'si-Albert model (adding one and two edges at a time). 100 simulations of networks grown to 525 nodes are averaged. The integrated specialization model and random specialization perform virtually the same, with little variance in both models. The BA model adding one edge doesn't achieve as high of a percentage, and exhibits more variance between simulations. For the BA model adding two edges, there is no consistent creation of non-trivial equitable partitions.}
         \label{fig:Compare_EP}
     \end{subfigure}
\end{figure}

For the integrated specialization model, we achieve 100\% non-triviality quite rapidly, as seen in Figure \ref{fig:EP}, with little standard deviation. We see similar results for repeated random minimal specialization (Figure \ref{fig:Compare_EP}). (As far as the author's know, there is no theory describing the occurrence of equitable partitions in BA models.) We can see that, on average, the BA network adding one node and edge evolves to have roughly 70\% non-triviality, with higher variance than other models. For BA networks grown by adding one node and two edges, there is no consistent creation of equitable partitions (Figure \ref{fig:Compare_EP}). Our model is potentially useful in the sense that it can be used to evolve a network to any percentage of non-triviality, given that the percentage is in the of the form $\frac{k}{n}$ for $ 0 \leq k \leq n$ where $n$ is the size of the starting network.

\section{Acknowledgements}
B. W. was partially supported by the NSF grant \#2205837 in Applied Mathematics. A. K. was partially supported by the NSF Graduate Research Fellowship Program under Grant \#2238458. Any opinions, findings, and conclusions or recommendations expressed in this material are those of the author(s) and do not necessarily reflect the views of the National Science Foundation.

\bibliographystyle{unsrt}

\bibliography{main}

 \end{document}